\documentstyle[aps,preprint,floats,prc]{revtex}

\begin{document}

     % BASIC DEFINITIONs
   \newcommand{\newc}{\newcommand}

 % GLOSSARY

    \newc{\ncf}{nucleon correlation function}
    \newc{\qcdsr}{QCD sum rules}
    \newc{\m}{\mbox{ }}

 % SYMBOLs

    %SYMMETRY

    \newc{\chb}{{\!\not\!C}}
    \newc{\pab}{{\!\not\!P}}
    \newc{\tib}{{\!\not\!T}}

    %LORENTZ TENSORs

    \newc{\gax}{\gamma_5}
      
    %vector
                                                                     
  \newc{\vpal}{p_{\alpha}}  \newc{\vqal}{q_{\alpha}}  \newc{\vlal}{l_{\alpha}}
  \newc{\vpbe}{p_{\beta}}   \newc{\vqbe}{q_{\beta}}   \newc{\vlbe}{l_{\beta}}
  \newc{\vpmu}{p_{\mu}}     \newc{\vqmu}{q_{\mu}}     \newc{\vlmu}{l_{\mu}}
  \newc{\vpnu}{p_{\nu}}     \newc{\vqnu}{q_{\nu}}     \newc{\vlnu}{l_{\nu}}
  \newc{\vpxi}{p_{\xi}}     \newc{\vqxi}{q_{\xi}}     \newc{\vlxi}{l_{\xi}}
  \newc{\vpet}{p_{\eta}}    \newc{\vqet}{q_{\eta}}    \newc{\vlet}{l_{\eta}}
  \newc{\vplm}{p_{\lambda}} \newc{\vqlm}{q_{\lambda}} \newc{\vllm}{l_{\lambda}}
  \newc{\vprh}{p_{\rho}}    \newc{\vqrh}{q_{\rho}}    \newc{\vlrh}{l_{\rho}}

  \newc{\gaal}{\gamma_{\alpha}}                                        
  \newc{\gabe}{\gamma_{\beta}}   
  \newc{\gamu}{\gamma_{\mu}}                                                    
  \newc{\ganu}{\gamma_{\nu}}     
  \newc{\gaxi}{\gamma_{\xi}}                                                    
  \newc{\gaet}{\gamma_{\eta}}    
  \newc{\galm}{\gamma_{\lambda}}                                               
  \newc{\garh}{\gamma_{\rho}}    

    %covector
                
  \newc{\cpal}{p^{\alpha}}  \newc{\cqal}{q^{\alpha}}  \newc{\clal}{l^{\alpha}}
  \newc{\cpbe}{p^{\beta}}   \newc{\cqbe}{q^{\beta}}   \newc{\clbe}{l^{\beta}}
  \newc{\cpmu}{p^{\mu}}     \newc{\cqmu}{q^{\mu}}     \newc{\clmu}{l^{\mu}}
  \newc{\cpnu}{p^{\nu}}     \newc{\cqnu}{q^{\nu}}     \newc{\clnu}{l^{\nu}}
  \newc{\cpxi}{p^{\xi}}     \newc{\cqxi}{q^{\xi}}     \newc{\clxi}{l^{\xi}}
  \newc{\cpet}{p^{\eta}}    \newc{\cqet}{q^{\eta}}    \newc{\clet}{l^{\eta}}
  \newc{\cplm}{p^{\lambda}} \newc{\cqlm}{q^{\lambda}} \newc{\cllm}{l^{\lambda}}
  \newc{\cprh}{p^{\rho}}    \newc{\cqrh}{q^{\rho}}    \newc{\clrh}{l^{\rho}}

  \newc{\haal}{\gamma^{\alpha}}
  \newc{\habe}{\gamma^{\beta}}
  \newc{\hamu}{\gamma^{\mu}}
  \newc{\hanu}{\gamma^{\nu}}
  \newc{\haxi}{\gamma^{\xi}}
  \newc{\haet}{\gamma^{\eta}}
  \newc{\halm}{\gamma^{\lambda}}
  \newc{\harh}{\gamma^{\rho}}

    \newc{\tena}[2]{   \sigma_{{#1}{#2}}            }
    \newc{\tenb}[2]{   \sigma_{{#1}{#2}}       \gax }
    \newc{\tenc}[2]{   {#1} {#2} - {#2} {#1}        }
    \newc{\tend}[2]{ ( {#1} {#2} - {#2} {#1} ) \gax }
    \newc{\tene}[5]{ \epsilon_{ {#1}{#2}{#3}{#4} } {#5}^{#3} \gamma^{#4}      }
    \newc{\tenf}[5]{ \epsilon_{ {#1}{#2}{#3}{#4} } {#5}^{#3} \gamma^{#4} \gax }
    \newc{\teng}[3]{ \mathaccent 94 {#3} ( {#3}_{#1} \gamma_{#2} - {#3}_{#2} \gamma_{#1} ) }       
    \newc{\tenh}[3]{ \mathaccent 94 {#3} ( {#3}_{#1} \gamma_{#2} - {#3}_{#2} \gamma_{#1} ) \gamma_5}

  % NUCLEON

    %NEUTRON                                  %PROTON

    \newc{\nmass}{M_{n}}                      \newc{\pmass}{M_{p}}
    \newc{\resn}{\lambda_{n}}                 \newc{\resp}{\lambda_{p}}
    \newc{\alphan}{\alpha_{n}}                \newc{\alphap}{\alpha_{p}}

    \newc{\ncon}{s_{0}^{n}}                   \newc{\pcon}{s_{0}^{p}}
    \newc{\betan}{\beta_{n}}                  \newc{\betap}{\beta_{p}}

    \newc{\ncharge}{Q_{n}}                    \newc{\pcharge}{Q_{p}}
    \newc{\namm}{{\mu}_{n}^{a}}               \newc{\pamm}{{\mu}_{p}^{a}}
    \newc{\nedm}{d_{n}}                       \newc{\pedm}{d_{p}}
    \newc{\nmm}{{\mu}_{n}}                    \newc{\pmm}{{\mu}_{p}}
                        
    \newc{\nfc}{F_{1}^{n}(0)}                 \newc{\pfc}{F_{1}^{p}(0)}
    \newc{\nfm}{F_{2}^{n}(0)}                 \newc{\pfm}{F_{2}^{p}(0)}
    \newc{\nfe}{F_{3}^{n}(0)}                 \newc{\pfe}{F_{3}^{p}(0)}
    \newc{\nfa}{F_{4}^{n}(0)}                 \newc{\pfa}{F_{4}^{p}(0)}

  % QUARK MASS, CHIRAL RADIUS, CHARGE

    \newc{\massq}{m_q}    \newc{\raq}{R_q}    \newc{\chq}{e_q}
    \newc{\massu}{m_u}    \newc{\rau}{R_u}    \newc{\chu}{e_u}
    \newc{\massd}{m_d}    \newc{\rad}{R_d}    \newc{\chd}{e_d}
    \newc{\masss}{m_s}    \newc{\ras}{R_s}    \newc{\chs}{e_s}
    \newc{\massQ}{M_Q}    \newc{\raQ}{R_Q}    \newc{\chQ}{e_Q}

  % CHIRAL PHASEs

    %theta bar
    \newc{\thetabar} {   { \bar{\theta} }     }
    \newc{\thetabarq}{ { { \bar{\theta} }_q } }
    \newc{\thetabaru}{ { { \bar{\theta} }_u } }
    \newc{\thetabard}{ { { \bar{\theta} }_d } }
    \newc{\thetabars}{ { { \bar{\theta} }_s } }
    \newc{\thetabarQ}{ { { \bar{\theta} }_Q } }

    %theta glue
    \newc{\thetag} { { \theta_G }    } 
    \newc{\thetagq}{ { \theta_{Gq} } }
    \newc{\thetagu}{ { \theta_{Gu} } }
    \newc{\thetagd}{ { \theta_{Gd} } }
    \newc{\thetags}{ { \theta_{Gs} } }
    \newc{\thetagQ}{ { \theta_{GQ} } }

    %theta quark
    \newc{\thetaq}{ {\theta _ q} }
    \newc{\thetau}{ {\theta _ u} }
    \newc{\thetad}{ {\theta _ d} }
    \newc{\thetas}{ {\theta _ s} }
    \newc{\thetaQ}{ {\theta _ Q} }

    %condensate
    \newc{\cond}[1]{ {\langle {#1} \rangle} _ {\thetaq,\thetag} }
    \newc{\varcond}[3]{ {\langle {#1} \rangle} _ {#2,#3} }

  % NCF NOTATIONs

    \newc{\pinu}{\Pi^{N}(p)}
    \newc{\pizero}{\Pi_{0}^{N}(p)}
    \newc{\pimunu}{\Pi_{\mu\nu}^{N}(p)}
   
    \newc{\npi}{\Pi^{n}(p)}
    \newc{\npizero}{\Pi_{0}^{n}(p)}
    \newc{\npimunu}{\Pi_{\mu\nu}^{n}(p)}

    \newc{\ppi}{\Pi^{p}(p)}
    \newc{\ppizero}{\Pi_{0}^{p}(p)}
    \newc{\ppimunu}{\Pi_{\mu\nu}^{p}(p)}
 
 % MATH FORMULA
    \newc{\deb}{\!\not\!D}
    \newc{\ftr}[2]{     \int d^{2\omega} {#2} \m e^{ i {#1} {#2} }     }

     \draft

     \preprint{\vbox{\it  \null\hfill\rm DOE/ER/41014-11-N97}\\\\}

     \title{On the Symmetry Constraints of CP Violations in QCD}
    
     \author{Chuan-Tsung Chan\thanks{E-mail address: 
             chan@alpher.npl.washington.edu}}

     \address{Department of Physics, Box 351560,
              University of Washington,
              Seattle, WA 98195-1560, USA}

     \date{\today}

     \maketitle

   \begin{abstract}

     %Abstract

Three symmetry constraints on the CP violations in QCD are discussed in this
paper.
In order to generate CP violating observables from QCD, these constraints 
require: (1) spontaneous chiral symmetry breaking, (2) explicit chiral 
symmetry breaking, e.g., finite quark masses, (3) $U_A(1)$ anomaly, in addition
to a nonzero $\thetabar$ parameter.
A pictorial illustration is used to unify and elucidate these constraints and
indicate a dual relation between quark mass and the quark condensate.
Based on the symmetry constraints, a dynamical suppression scenario to solve 
the strong CP problem within QCD is examined. 
We conclude that a solution of the strong CP problem has to involve physics
beyond the standard model.

   \end{abstract}


     \narrowtext

   \section{Motivations}
     \label{sec:z2}

     %Motivations

In spite of many research efforts, CP violation in the subnuclear world remains
one of the most important challenges yet to be solved by physicists 
\cite{CP:review}.
While CP violations in the underlying dynamics of elementary particles are 
required to explain the observed baryon asymmetry in our universe \cite{BA:SA},
we have only one system ( $K_L, K_S$ mesons, in their decay modes into two or 
three pions ) which demonstrates that CP is not a good symmetry of the weak 
interactions \cite{CP:K meson}. 
The experimental evidence we have acquired is insufficient to pin down a unique 
theory.
                   
On the other hand, it is believed that we have a good understanding of parity 
violation in the weak interactions, and the existing measurements indicate 
that strong and electromagnetic interactions conserve C, P and T to a high 
accuracy, e.g., the measurements of a neutron electric dipole moment 
\cite{NEDM:CN exp}.
While the above interactions are successfully described by the standard model,
it is possible to incorporate CP violation in the $K$ meson system and thus 
provide predictions for other phenomena, e.g., CP violation in the $B$ meson
system, and/or explanations of the baryon asymmetry. 
In so doing, one is naturally led to the following question:
Why are the existing discrete symmetry breakings only observed in the weak 
sector and not in the strong one? 
Such an innocent puzzle may have profound ramifications; for example, if one 
insists that the strong interaction is governed by a non-Abelian gauge theory 
like QCD, the nontrivial topological structure ( $\grave{a} \m  la$ instantons 
) implies a $\theta$ vacuum which violates CP \cite{QCD:theta}. 
The situation is further complicated by the presence of a quark mass
matrix, which connects weak CP violations to the strong ones under chiral 
transformations.
From a theoretical perspective, it is not clear why QCD should conserve CP and 
this is generally referred to as a strong CP problem \cite{QCD:CP}.

One can answer this question by asserting that the parameter ( which will be 
specified later ) associated with CP violations in QCD is small, so that the 
observed effects are tiny. 
We shall consider such a solution undesirable, as a vanishing parameter without
increasing the symmetry of QCD should be considered "unnatural", according to 
t'Hooft \cite{FT:natural}. 
Furthermore, there is no way to calculate the CP violating parameter from QCD
itself.
However, if we treat QCD as a low-energy effective interaction of some
high-energy theory ( e.g., some grand unified theory ), then such a CP 
violating parameter can be calculated and can be shown to be naturally small in
certain models \cite{CP:A Nelson}. 
Another solution to the strong CP problem is given by introducing an extra 
pseudoscalar particle, namely, the axion \cite{CP:PQ}.
In either case, an explanation of the small CP violating parameter necessarily
involves physics beyond the standard model.
Nevertheless, these scenarios seem to receive little experimental support at 
this moment.

In between these alternative solutions to the strong CP problem, one can 
imagine that there exists the possibility that CP violations in a non-Abelian
gauge theory are dynamically suppressed.
That is, the intrinsic symmetry of the theory leads to small coefficients of CP
violating observables, e.g., neutron electric dipole moment,
in company with a function of the CP violating parameter which can be of order
$O(1)$. If this were the case, a natural solution of CP violations in QCD can be
obtained without invoking extra particles or modifications of the gauge 
structure.
It is to this point that this paper is addressed, and we shall examine a 
dynamical suppression mechanism in terms of symmetry constraints on CP 
violations in QCD, which is the main subject of this paper.

   \section{Introduction}
     \label{sec:z3}

     %Introduction

Even though we shall focus on the CP violations of QCD in this paper, it is
helpful to begin our discussion with CP violations in the standard model, 
which includes the electroweak theory and QCD as two ingredients. 
As the definition of the strong CP problem is a subtle issue, it is important
to specify our domain in order to avoid extra complications.

First of all, by standard model we mean the minimal theory with a $SU_C (3) 
\times SU_L (2) \times U(1)$ gauge structure and one Higgs doublet together 
with fundamental fermions ( quarks and leptons ) as matter fields. 
Second, there are three places in the standard model Lagrangian where
one can naturally incorporate CP violation without modifying the gauge 
structure and/or changing the particle content of the theory. These are: ($1$) 
charged current vertices; ($2$) Yukawa couplings between fermions and Higgs 
particle; ($3$) gauge field anomalies. 
Third, it is important to notice that the three sources of CP violation are not
independent, due to the reparametrization invariance of the generating
functional. 
In particular, under a general chiral rotation in the fermion flavor space, 
it is possible to shift the CP violating parameters among the three terms. 
Consequently, there are only two independent CP violating parameters in the 
three-generation standard model with massless neutrinos. 
Fourth, if we are only concerned with low energy physics, it is useful to look
at the effective theory, where we integrate out all heavy particles ( e.g., 
Higgs particle, $W$ and $Z$ bosons, plus heavy fermions ) of the standard 
model. In the low energy effective theory, the first two terms mentioned above
reduce to: ($1'$) flavor changing 4-fermion ( weak ) interactions and 
($2'$) fermion mass matrix, and the CP violation can be characterized by a
Cabibbo-Kobayashi-Maskawa ( CKM ) phase\footnote{The existence of such a CP 
violating phase requires at least three generations of fermion doublets.}. 
If we negelect the electroweak interactions in our discussion, as we shall do 
later, the combined effects of ($2'$) and ($3$) can be characterized by a 
single parameter, which is referred to as the $\thetabar$ parameter.
Finally, while it is complicated to give reparametrization and renormalization
group invariant definitions of these two CP violating parameters in the 
three-generation standard model, it is worthwhile to notice that there exists a
particular representation of the standard model Lagrangian in which ($2'$) is
CP even but ($1'$) and ($3$) violate CP.
Henceforth, we shall refer to the effects derived from the CKM phase as weak CP
violations and those related to the $\thetabar$ parameter as strong CP 
violations\footnote{The $U(1)$ gauge anomaly is physically irrelevant because 
of the trivial topological structure of the Abelian gauge theory. 
The $SU_L (2)$ gauge anomaly, due to the $V-A$ structure of weak interaction, 
is also unimportant in our discussion.}.

   \section{Strong CP Violations in QCD}
     \label{sec:z4}

     %Strong CP violations in QCD

Having defined the strong CP problem within the framework of the standard
model, we shall write down the explicit definitions to set up proper notations
for further reference. For the sake of simplicity, we shall discuss a theory
with one quark flavor\footnote{The simplification is due to the fact that
strong CP violation is a flavor-singlet problem.}. The generalization to a
multi-flavor case can be found elsewhere \cite{QCD:C-T Chan}.

Normally, the ( CP conserving ) QCD Lagrangian is given by
\begin{eqnarray}
       { \cal L }_{QCD} &\equiv&           \bar{\psi} i \deb \psi
                                  + \massq \bar{\psi}        \psi
                                  + \frac{1}{4}  G^2               \\
     \mbox{where} \hspace{2cm}
                  \deb  &\equiv& ( \m \partial_\mu
                                    + i g_s  B_\mu^a \frac{ \lambda^a }{2} \m )
                                  \cdot \gamma^\mu
\end{eqnarray}
   The meanings of various symbols are:
      \begin{eqnarray*}
      \psi                  &:& \mbox{quark field }                          \\
      \bar{\psi}            &:& \mbox{Dirac adjoint of the quark field, }
                                \bar{\psi} \equiv {\psi}^{\dagger} \gamma_0  \\
      B_\mu^a               &:& \mbox{gluon field, } a = 1,..,8              \\
      \frac{\lambda^a}{2}   &:& \mbox{generators of the color } SU(3)
                                \mbox{ gauge group, } a = 1,..,8             \\
      G_{\mu\nu}            &:& \mbox{gluonic tensor field, } G_{\mu\nu} \equiv
                               [\m \partial_\mu + i g_s B_\mu \m,
                                \m \partial_\nu + i g_s B_\nu \m],           \\
                            & & B_\mu \equiv B_\mu^a \frac{\lambda^a}{2},
                                \m \m G^2 \equiv  G_{\mu\nu} G^{\mu\nu}      \\
      g_s                   &:& \mbox{strong coupling constant in QCD}         
     \end{eqnarray*}
 
       To calculate various correlation functions in the quantum theory, it is
     useful to define the QCD generating functional ( denoted by Z ):
     \begin{equation}
      Z[ \m \zeta, \bar{\zeta}, J_\mu \m ] \equiv
         \frac{1}{N} \int [ D \psi ] [ D \bar{\psi} ] [ DB_\mu ]
        e^ { iS_{QCD} + \bar{\zeta} \psi + \bar{\psi} \zeta + J_\mu B^\mu }
     \label{eq:gen}
     \end{equation}
     where the QCD action 
     \begin{equation}
      S_{QCD} \equiv \int d^4 x \m { \cal L }_{QCD}
     \end{equation}

      The normalization constant for the generating functional is
     \begin{equation}
      N \equiv  \int [D\psi] [D\bar{\psi}] [DB_{\mu}] e^ {iS_{QCD}}
     \end{equation}
      such that
     \begin{equation}
      Z[ \m \zeta=0, \m \bar{\zeta}=0, \m J_\mu=0 \m ] = 1
     \end{equation}

    Notice that in the generating functional ( Eq.~\ref{eq:gen} ), the fermion 
    field $\psi, \bar{\psi}$ and the gluon field $B_\mu^a$ are dummy 
    variables; only the external source fields $\zeta, \bar{\zeta}, J_\mu$ 
    specify the physical ground state ( QCD vacuum ) of the theory. 
    Therefore, we can redefine these dummy variables freely without changing
    the physical contents of the theory. 
    In particular, we can perform a $U_A (1)$ chiral rotation on the fermion 
    field:
    \begin{eqnarray}
      \psi \rightarrow { \psi '} &\equiv& e^{i \theta \gamma_5} \psi
      \label{eq:chiral} \\
      \mbox{or \hspace{2cm} } { \psi '}_i &=&
      [\m  \cos \theta ( I )_{ij}
      + i  \sin \theta ( {\gamma_5} )_{ij} \m ] \m {\psi}_j
    \end{eqnarray}

    While it is clear that the quark mass term  $\m {\massq} \bar{\psi} \psi \m$
    transforms into $\m \massq \bar{\psi} e^{ 2 i \theta \gamma_5 } \psi \m$
    under a $U_A (1)$ chiral rotation, it is not a trivial task to show that
    a $U_A (1)$ chiral rotation is not an unitary transformation ( $U_A (1)$
    anomaly ) and a Jacobian associated with this change of variables in the 
    functional space has to be implemented. 
    As shown by Fujikawa \cite{QCD:Fuji}, the fermionic measure in the
    functional integral
    $[ D \psi ] [ D \bar{\psi} ]$ transforms, under a $U_A (1)$ chiral
    rotation,
    \begin{equation}
     [ D \psi ] [ D \bar{\psi} ] \rightarrow [ D \psi ] [ D \bar{\psi} ]
      \m e^{ i \frac{ g_s^2 2 \theta }{ 32 \pi^2 } \int d^4 x G \tilde G
              ( x ) }
    \end{equation}

    Hence, we generate a new ( but equivalent ) QCD Lagrangian with two extra
    terms ( together with a change of the chiral phases of the external fermion
    source fields $\zeta$, $\bar \zeta$ ):
    \begin{equation}
     \massq \m \bar{\psi}   \m ( e^{i 2 \theta \gamma_5} - 1 ) \m \psi \m 
                            \approx 
   i \massq \m \sin 2 \theta \m \bar{\psi} \gamma_5 \psi,  \hspace{1.5cm}
     \frac{ g_s^2 2 \theta }{ 32 \pi^2 } G \tilde G  \nonumber
    \end{equation}

    Several comments are in order:

    (1) These two terms carry the same quantum numbers and both are odd under
    parity ( P ) and time reversal ( T ) transformations. We shall refer to the
    former as a quark pseudomass term, and the latter as a gluon anomaly term. 
    We emphasize that these are the lowest dimensional CP violating operators
    that one can write down in the QCD Lagrangian, consistent with basic 
    requirements\footnote{These requirements include: (1)
    Hermiticity, (2) Lorentz invariance, and (3) gauge invariance.} of a 
    relativistic quantum field theory.

    (2) We can generalize the discussion to a ( generally CP violating ) QCD
    Lagrangian with an arbitrary quark pseudomass term $ \massq \bar{\psi} 
    e^{i \thetaq \gamma_5} \psi$ and a gluon anomaly term $\frac{ g_s^2 
    \thetag }{ 32 \pi^2 } G \tilde G$.
    \begin{equation}
       { \cal L }_{QCD; \thetag, \thetaq} \equiv 
                                        \bar{\psi} i \deb                 \psi
                               + \massq \bar{\psi} e^{i \thetaq \gamma_5} \psi
                               + \frac{1}{4}  G^2 
                               + \frac{ g_s^2\thetag }{ 32 \pi^2 } G \tilde G
    \end{equation}     
    It can be shown that, under a $U_A (1)$ chiral rotation
    ( Eq.~\ref{eq:chiral} ), both $\thetaq$ and $\thetag$ change by $2 \theta$.
    Therefore, the difference
    \begin{equation}
    \thetabar \equiv \thetag - \thetaq
    \end{equation}
    is an invariant of the $U_A (1)$ chiral rotation, which can be used to
    label the classes of equivalent QCD Lagrangians. Since the physical
    observables are independent of the representations of the generating
    functional, we conclude that any CP violating observable has to be 
    proportional to the $U_A (1)$ invariant chiral phase $\thetabar$.

    (3) The $\thetabar$ parameter, being a difference between two chiral phases,
    is an angular variable with period $2 \pi$ ( in the case of one quark
    flavor ). Consequently, any physical observable has to be a periodic
    function of the $\thetabar$ parameter.

    (4) In the muti-flavor case of QCD, the number of quark chiral phases is
    equal to the number of quark flavors. Nevertheless, since we can perform a
    $U_A (1)$ chiral rotation independently on each flavor, there is still only
    one physical parameter which characterizes the strength of strong CP 
    violations.
    In that case, the $\thetabar$ parameter is defined as 
    \begin{equation}
    \thetabar \equiv \thetag - \sum_j \thetaq_j
    \end{equation}
    where $j$ is the flavor index for light quarks.

   \section{Symmetry Constraints of CP Violations in QCD}
     \label{sec:z5}
	
     %Symmetry constraints of CP violations in QCD 
                                      
 The previous discussions seem to suggest that there are close relationships 
 between the $U_A(1)$ chiral symmetry and the CP violations in QCD. 
 Indeed, it is necessary that the chiral symmetry is broken both spontaneously 
 and explicitly so that strong CP violations are possible. 
 We shall refer to these relations as symmetry constraints and discuss their 
 meanings and implications in this section.

  (1) Non--Perturbative Nature of CP Violations in QCD

   Given the fact that the gluon anomaly term $G \tilde G$ can be
   written as a total divergence of the Chern-Simon current $K_{\mu}$,
   \begin{eqnarray}
                     K_\mu &\equiv&
                     \frac{g_s^2}{16 \pi^2} \sum \epsilon_{\mu \nu \rho \sigma}
                     B^{a \nu} \{ \partial^\rho B^{a \sigma} +
                     \frac{1}{3}   f_{abc}      B^{b \rho}   B^{c \sigma} \} \\
                     \partial^\mu K_\mu &=& \frac{g_s^2}{32 \pi^2} G \tilde G
   \end  {eqnarray}
    it is not too surprising that this term has no effect on a perturbative
    calculation of any CP violating observable. 
    Since a potential modification of the perturbative expansion caused by the
    gluon anomaly term can only be related to the gluonic propagator in the 
    Feynman rules, it turns out such a contribution to the gluonic propagator
    is zero, as can be verified by a direct calculation.

    Since we can always perform chiral rotations to shift the strong CP
    violating phases into the gluon anomaly term ( i.e., $\thetaq=0, \m 
    \thetabar=\thetag$ ), and any physical observable should not depend on the 
    particular representation we choose to do a calculation, we conclude that
    the strong CP violation has to be a purely nonperturbative effect. 
    However, it remains to see how a calculation of CP violating observables 
    based on the quark pseudomass term leads to the same conclusion, if we 
    insist on the reparametrization invariance of the CP violating observables.
    This is a problem, because it seems that the presence of a quark pseudomass
    term will lead to a modification of the quark propagator, hence generate 
    contributions to the CP violating observables in perturbative calculations.
    Apparently, such an effect is in contradiction to our previous
    observation.

    Indeed, if we perform a calculation of a quark electric dipole moment (
    denoted as qEDM ) using a perturbative expansion on both fine structure 
    constant ${e^2}/{\hbar c}$ and strong coupling constant $g_s$, we do find a
    finite contribution to a tensor structure which corresponds to a qEDM, with
    the strength proportional to the quark chiral phase $\sin \thetaq$, see 
    Fig.~\ref{fig:Quark EM moments}.
    Nevertheless, we should be careful not to interpret this result as a 
    physical observable. As we discussed before, $\thetaq$, by itself,
    is a representation-dependent parameter and CP violating physical
    observables should depend only on $\thetabar$. What goes wrong here?

    The answer is that there are other representation-dependent chiral phases
    we should include in the extraction of a physical observable associated
    with a chirally covariant tensor. 
    For example, in the case of a particle EDM, the relevant tensor 
    $i \sigma_{\mu \nu } \gamma_5$ is mixed with the tensor associated with the 
    anomalous magnetic moment ( denoted as AMM ) $\sigma_{\mu \nu}$ under a 
    $U_A (1)$ chiral rotation on the particle field. 
    Besides that, the same $U_A(1)$ chiral rotation also causes the mass of the
    particle to develope a chiral phase, 
    $M \rightarrow M e^{i \alpha \gamma_5}$, which can be viewed as a mixing 
    between $I$ and $i \gamma_5$ tensors, see Fig.~\ref{fig:Quark propagator}. 
    We need to subtract the relative phases, $ \arctan ( \m \frac { \mbox{EDM} }
    { \mbox{AMM} } \m )$ and $\alpha$, in order to define a 
    representation-independent answer for the physical observables. 
    It can be verified that in the calculation of the qEDM, both $ \arctan ( \m
    \frac { \mbox{EDM} }{ \mbox{AMM} } \m ) $ and $\alpha$ are
    equal to $\thetaq$; hence there is no quark EDM from the perturbative
    calculation in any representation of the QCD generating
    functional\footnote{This point has been emphasized by E.P. Shabalin 
    \cite{CP:Shabalin}. However, the argument in support of the conclusion in
    these works seems unclear to the present author.}.

    With a suitable generalization, one can convince oneself that this
    conclusion holds true to all orders in QCD, i.e., the inclusions of any
    higher order loops does not affect the conclusion.  
    One can also apply the same argument to the bound states of quarks, i.e.,
    hadrons, and show that any CP violating observables have to come from the 
    nonperturbative contributions of QCD\footnote{By perturbative 
    contributions, we mean those arising from an expansion of any physical 
    observable in a power series of the strong coupling constant $g_s$.
    A purely nonperturbative observable has zero coefficients to all order in 
    its power series expansion. For example, $\langle \bar q q \rangle \propto
    e^{ - 1/{g^2_s} }$.}.

    Since the nonperturbative property of QCD can be characterized in terms of
    the spontaneous chiral symmetry breaking, which is manifested by the
    presence of vacuum condensates, we can rephrase the conclusion in the
    following way:

    If there is no spontaneous chiral symmetry breaking in QCD, there is no
    strong CP violation, even with a nonzero $\thetabar$.

  (2) Chiral Limit and CP Violations in QCD

    Another important constraint on the strong CP violations has to do with a
    nonvanishing quark mass term in the QCD Lagrangian. It was first pointed
    out by Peccei and Quinn \cite{CP:PQ} that there is no strong CP violation 
    if there exists a massless quark in the QCD Lagrangian.

    This constraint can be understood in terms of the functional integral 
    formalism.
    Since a quark chiral phase $\thetaq$ is ill-defined if the quark mass is
    zero, we can take advantage of this fact to rotate away ( via an $U_A (1)$
    chiral transformation ) the gluon chiral phase $\thetag$, so that any QCD
    Lagrangian with a massless quark is equivalent to a CP conserving one.
    Therefore, it is necessary to have all quark masses finite to generate a 
    CP violating observable from QCD.

    The point of this argument is that we need to have a well-defined chirally
    covariant phase, in addition to the gluon chiral phase $\thetag$, such 
    that, after the reduction of the irrelevant degrees of freedom by 
    reparametrization invariance, we are left with a chirally invariant CP 
    violating parameter, e.g., $\thetabar$. 
    For instance, we can replace the quark mass term by a higher dimensional 
    operator which violates the chiral symmetry explicitly,
    e.g., $\bar q \sigma_{\mu \nu} e^{i \beta \gamma_5} q G^{\mu \nu}$
    \cite{NEDM:KW}. 
    In this case, even though the quark is massless, we still can have a CP
    violating observable; proportional to the chirally invariant phase 
    $\thetag - \beta$.
    Consequently, the second constraint can be phrased as follows:
    
    If there is no explicit chiral symmetry breaking, there is no strong CP
    violation in QCD.

  (3) $U_{A}(1)$ Anomaly Constraint of CP Violations in QCD

    The third constraint was first discussed by M.A. Shifman, A.I. Vainshtein 
    and  V.I. Zakharov \cite{CP:SVZ}, and then rediscovered by S. Aoki, 
    A. Gocksch,  A.V. Manohar and S.R. Sharpe \cite{CP:Aoki}.
       They pointed out that, in a diagrammatical language, the strong CP 
       violations only contribute to physical observables through the 
       internal fermion loops with a pseudo--mass insertion\footnote{Therefore,
       it is impossible to calculate the effect of strong CP violations in a
       quenched lattice calculation \cite{CP:Aoki}.}.
       Such a diagram is a manifestation of the anomalous Ward
       identity associated with the flavor singlet axial current and has a 
       close relationship with the $U_{A}(1)$ anomaly in QCD.
       This connection has been examined in the context of chiral perturbation
       theory by S. Aoki and T. Hatsuda \cite{CP:A&H}, and H-Y Cheng 
       \cite{CP:H-Y Cheng}.
    Essentially, this constraint requires that the chiral anomaly provides a 
    solution to the $U_A(1)$ problem \cite{CP:SVZ} \cite{UA1:tHooft}.
       If this is not the case, then there is no strong CP violation.
       In a hadronic calculation, this constraint implies that CP violating 
       observables should be proportional to the difference between $m_\pi^2$
       and $m_{\eta'}^2$ \cite{CP:A&H} \cite{CP:H-Y Cheng}. 
       A quantitative realization of this constraint in QCD implies that CP 
       violating observables should be proportional to the anomalous gluon 
       condensate $\cond{G \tilde G}$ \cite{NEDM:C-T Chan}.

   It should come as no surprise that these three constraints are not
   independent since ultimately both the anomalous gluon condensate 
   $\cond{G \tilde G}$ and the quark chiral radius \cite{QCD:C-T Chan}
   \begin{equation}
    R_q^2 \equiv { \left[ \m   \cond{ q          \bar q } \m \right] }^2 +
                 { \left[ \m i \cond{ q \gamma_5 \bar q } \m \right] }^2 
   \end{equation}
   can be related to the QCD scale $\Lambda_{QCD}$. 
   Indeed, through the use of a generalized anomalous Ward identity, we can 
   prove that $\cond{G \tilde G}$ is propotional to the product of $m_q, R_q$
   and $\sin \thetabar$ \cite{QCD:C-T Chan}.

   To summarize, the bottom line of the study of symmetry constraints is that
   we need three finite QCD parameters:
   $m_q, R_q$ and $\sin \thetabar$ to generate a CP violating observable from
   QCD.

   \section{A Graphic Illustration of the Symmetry Constraints of 
            CP Violations in QCD}
     \label{sec:z6}

     %A Graphic Illustration of the Symmetry Constraints of 
%CP Violations in QCD

From the previous discussions, it is clear that to generate a CP violating 
observable from QCD, we need three nonzero parameters, $m_q, R_q$ and 
$\thetabar$. 
This conclusion was derived through the use of a functional integral formalism.
While these elegant derivations are exact and nonperturbative in nature, they
lack the intuition and simplicity to help us grasp the basic ideas.
In view of this, we would like to show a different approach to understand how 
these symmetry constraints work in QCD.

In fact, there is a simple way to visualize the symmetry constraints without
relying on the functional integral formalism. Specifically, we shall use a
graphic illustration to show that there is no strong CP violation if there is
no spontaneous chiral symmetry breaking and/or the quark mass 
vanishes\footnote{As we have explained before, the $U_{A}(1)$ anomaly 
constraint is not really independent from the first two constraints. Hence, we
shall ignore this constraint in this section.}.

    To begin with, the set of CP violating QCD Lagrangians with two CP
    violating parameters, as defined in Sec.~\ref{sec:z4}, can be 
    represented in a two dimensional plane ( phase space ),
    where a given pair of CP violating parameters
    corresponds to a point on the plane. It is useful to choose the polar
    representations for the CP violating parameters ( i.e., $\thetaq,\thetag$ )
    so that the periodic structures of these parameters implies an
    identification of
    the boundaries of the squares ( e.g., $\left[ \m \thetaq=0 \m     \right]
                                   \equiv  \left[ \m \thetaq=2 \pi \m \right]$ )
    and the phase space of the CP violating QCD Lagrangians becomes a torus,
    see Fig.~\ref{fig:CP torus}.

    On the two dimensional plane, we can identify the equivalent classes of
    the CP violating QCD Lagrangians ( those with the same $\thetabar$ ) by
    the straight lines $ \thetag - \thetaq = constant $. 
    After the identification of the boundaries of the square, these straight 
    lines map onto a family of nonintersecting closed loops winding over the 
    torus. 
    Any two points on the same curve, which correspond to equivalent CP 
    violating QCD Lagrangians with the same $\thetabar$, describe the same 
    physics ( reparametrization invariance ), see Fig.~\ref{fig:equiv class}.

    The connection with the symmetry constraints is established once we
    specify the length scales of the torus: the large radius, conjugate to the
    $\thetag$ angular variable, is related to some function of $R_q$,
    $ f( R_q ) $; and the small radius, conjugate to the $\thetaq$ angular 
    variable, is some function of $m_q$, $ g ( m_q ) $.
    Since we are only interested in a qualitative description of the symmetry
    constraints in this section, the actual form of the function is not 
    important, except that the function has to vanish when its argument is zero.
    With these specifications, we can study the change of geometries of the 
    torus in two special limits:

    (1) chiral limit ( $ m_q \rightarrow 0 $ ):

    In this limit, as the small radius $ g ( m_q ) $ shrinks to zero, the torus
    degenerates into a circle with radius $ f( R_q ) $ and all equivalent loops
    collapse onto the same circle. 
    If we insist on the single-valuedness of the physics as all the equivalent
    classes collapse onto the CP conserving one, it is natural to conclude
    that all strong CP violations vanish, see Fig.~\ref{fig:chiral limit}.

    (2) no spontaneous chiral symmetry breaking ( $ R_q \rightarrow 0$ ):

    In this limit, as the large radius $ f( R_q ) $ shrinks to zero, the torus
    degenerates into a sphere with radius $ g ( m_q ) $.
    The equivalent loops become eight--shaped curves and they all
    intersect with the CP conserving loop at two points.
    Again, using the single-valuedness argument, we conclude that if there is
    no spontaneous chiral symmetry breaking ( $ R_q = 0 $ ), the QCD
    Lagrangians conserve P and CP even with a nonzero $\thetabar$, see 
    Fig.~\ref{fig:second limit}.

    It is worth mentioning that such a graphic illustration indicates a
    dual relationship between $m_q$ and $R_q$.
    In addition, we hope that the geometrical pictures can be used 
    quantitatively. 
    For example, by choosing some suitable functions $ g ( m_q ) $ and 
    $ f( R_q ) $ of both radii of the torus, we might be able to relate the EM
    moments of particles to certain geometrical measures
    ( e.g., surface area enclosed by certain contour on the torus ) or fluxes 
    through the loops.

   \section{Why Strong CP Violations Are Small}
     \label{sec:z7}

     %Why Strong CP Violations Are Small

As we mentioned in the motivation, the discussions of the ( chiral ) symmetry
constraints on the strong CP violations in QCD is not just of theoretical
interest. 
One practical implication we hope to infer from these general constraints is 
whether QCD can cure the strong CP problem by itself without resorting to an 
unnaturally small $\thetabar$ parameter.
This suggestion might seem too ambitious in view of the amazing experimental
data on the CP violation observables, e.g., the search for a neutron electric
dipole moment at a level of $10^{-26} e \cdot cm$.
To achieve such a tiny effect ( the ratio of the EDM to the AMM of a neutron
is less than $10^{-12}$ \cite{NEDM:CN exp} ) requires a delicate cancellation in
any calculation if $\thetabar$ is of order unity.
However, we feel that it is worthwhile to consider this problem in a quantative
way, as the result involves only nonperturbative contributions,
which can be related to other observables in the physics of the strong
interactions.
Such a study has been done in the framework of the low-energy effective theory
of QCD \cite{CP:A&H} \cite{CP:H-Y Cheng}. 
A more direct approach, based on the QCD parameters, appeared only recently 
\cite{NEDM:C-T Chan}.

If we formulate the calculation of the CP violating observables based on the 
QCD Lagrangian, one useful technique is provided by the method of operator 
product expansion ( OPE ), as examplified in the practice of QCD sum rule 
calutions \cite{QSR:SVZ} and heavy quark expansion \cite{HQ: }.
This is a systematic expansion of physical observables in the dimensions of
QCD operators ( over some suitable energy scale, e.g., $\frac{\Lambda_{QCD}^2}
{Q^2}$, $\frac{\Lambda_{QCD}^2}{M^2}$ ).
The physics associated with short-distance and long-distance fluctuations are
factorized into Wilson coefficients and matrix elements of the QCD operators,
respectively.
Such a scheme is particularly useful in QCD because the nonperturbative aspects
of the theory are parametrized in terms of various condensates, for instance:
quark condensate $ \langle \bar q q \rangle $ and anomalous gluon condensate
$\cond{G \tilde G}$ \cite{QCD:C-T Chan}, and the nonperturbative contributions
to the Wilson coefficients become important only in higher dimensional operators
\cite{QSR:SVZ}. 
If we assume that low energy hadronic observables can be approximately 
saturated by the first few lower dimensional operators and the Wilson
coefficients are dominated by perturbative contributions, then OPE series of a
given correlation function can give us quantative information of low energy 
hadronic observables in terms of QCD parameters.

For a given physical quantity, the numerical value is a function of the Wilson
coefficients and various matrix elements ( condensates ), together with other
parameter, e.g., $\thetabar$.
If we can show that the Wilson coefficients are small ( e.g., due to small 
quark masses and/or other cancellations ) and the observable only receives 
contributions from higher dimensional operator because of the symmetry
constraints, then the smallness of that observable can be considered as 
natural, because such a small number comes from a dynamical suppression instead
of an unexplained tiny input parameter.
In the case of CP violating observables, we can establish a relation between 
hadronic observables and QCD parameters and thus use it to answer the question
of why the strong CP violations are small.

We have performed a calculation of the nucleon electric dipole moments, based 
on the method of QCD sum rule \cite{NEDM:C-T Chan}, to study the above
question.
While we are able to demonstrate that the three chiral symmetry constraints 
hold explicitly in our calculation without assuming a small $\thetabar$, the 
numerical result indicates that the dynamical suppression is not sufficient to
achieve the experimental upper bound for the neutron EDM\footnote{Our 
calculation, which generates a functional dependence of the nucleon EDM on the
$\thetabar$ parameter, satisfies the current upper limit from cold neutron 
experiments \cite{NEDM:CN exp}, with a $\thetabar$ parameter of the order 
$10^{-9}$.}.
Thus, the answer to the question whether QCD can cure the strong CP problem by
itself, from the perspective of the symmetry constraints, seems to be 
negative.

   \section{Summary and Conclusion}
     \label{sec:z8}

     %Summary and Conclusion

In this paper, we study the symmetry constraints on the strong CP violations in
QCD. Previous findings on the special features of the strong CP problem,
including: (1) its nonperturbative nature; (2) the 
importance of nonzero quark masses; (3) the chiral anomaly constraint, are
unified and elucidated in a dual relation between quark mass ( $m_q$ ) and the
quark condensate ( $R_q$ ), which can be visualized in a pictorial way without
much mathematical complication. 

We examine the possibility of a dynamical suppression mechanism of CP violating
observables, e.g., neutron electric dipole moment, due to the symmetry 
constraints in QCD. 
In a sum rule calculation of the nucleon EDMs \cite{NEDM:C-T Chan}, we obtain a
negative answer to the question of the existence of a natural solution to the 
strong CP problem within QCD. 
Hence, to explain a small $\thetabar$ parameter necessarily involves physics 
beyond the minimal standard model.

   \acknowledgments

     The author would like to thank Prof. E.M. Henley, Prof. T. Hatsuda, 
Prof. A. Nelson, Prof. P. Arnold and Dr. T. Meissner for many useful 
discussions.
This work was supported by Nuclear Theory Group of Department of Physics at
University of Washington, under the grant DE-FG-03-97ER41014.

      \newpage
  
 \begin{figure}
    \vspace{8cm}
    \caption{Quark propagator in the presence of pseudo-mass term}
    \label{fig:Quark propagator}
 \end{figure}

 \begin{figure}
    \vspace{12cm}
    \caption{Quark EM moments in the presence of pseudo-mass term}
    \label{fig:Quark EM moments}
 \end{figure}

 \newpage

 \begin{figure}
    \vspace{10cm}
    \caption{The phase plane and CP torus of QCD}
    \label{fig:CP torus}
 \end{figure}

 \begin{figure}
    \vspace{10cm}
    \caption{The equivalent classes of QCD Lagrangian}
    \label{fig:equiv class}
 \end{figure}

 \newpage

 \begin{figure}
    \vspace{10cm}
    \caption{The chiral limit of the CP torus of QCD}
    \label{fig:chiral limit}
 \end{figure}

 \begin{figure}
    \vspace{10cm}
    \caption{The chiral symmetric CP torus of QCD}
    \label{fig:second limit}
 \end{figure}


\begin{references}

     \bibitem{CP:review}
        {\it CP violations,\/} edited by C. Jarlskog
        ( World Scientific, Singapore; Teaneck, 1989 ).

        {\it CP violation,\/} edited by L. Wolfenstein
        ( North-Holland, Amsterdam; New York, 1989 ).

\bibitem{BA:SA}
       % VIOLATION OF CP INVARIANCE, C ASYMMETRY, AND BARYON ASYMMETRY
       % OF THE UNIVERSE. By A.D. Sakharov, 1967.
         A.D. Sakharov, 
         Pis'ma Zh. Eksp. Teor. Fiz. {\bf 5},  32  ( 1967 );
         JETP Lett.                  {\bf 5},  24  ( 1967 );
         Sov. Phys. Usp.             {\bf 34}, 392 ( 1991 ).
       % ( No.5 ).
       % ( Reprinted in *Kolb, E.W. (ed.), Turner, M.S. (ed.):
       % The early universe* 371-373,
       % and in *Lindley, D. (ed.) et al.:
       % Cosmology and particle physics* 106-109, 
       % [Usp. Fiz. Nauk 161 (1991) No. 5 61-64] )

\bibitem{CP:K meson}
         J.H. Christenson, J.W. Cronin, V.L. Fitch and R. Turlay,
       % Evidence for the 2pi decay of the K^0_2 meson,   
         Phys. Rev. Lett. {\bf 13}, 138 (1964).
  
\bibitem{NEDM:CN exp}
       % NEW MEASUREMENT OF THE ELECTRIC DIPOLE MOMENT OF THE NEUTRON.
         I.S. Altarev {\it et al.,\/} 
       % Yu.V. Borisov, N.V. Borovikova, S.N. Ivanov, E.A. Kolomenskii,
       % M.S. Lasakov, V.M. Lobashev, V.A. Nazarenko, A.N. Pirozhkov, 
       % A.P. Serebrov, Yu.V. Sobolev, E.V. Shulgina, A.I. Egorov 
       % ( St. Petersburg, INP & Moscow, INR ). 1992. 
         Phys. Lett. B {\bf 276}, 242 ( 1992 ).

       % A SEARCH FOR THE ELECTRIC DIPOLE MOMENT OF THE NEUTRON.
         K.F. Smith {\it et al.,\/}
       % N. Crampin, J.M. Pendlebury, D.J. Richardson, D. Shiers (Sussex U.),
       % K. Green, A.I. Kilvington, J. Moir, H.B. Prosper, D. Thompson 
       % (Rutherford), N.F. Ramsey (Harvard U.),
       % B.R. Heckel, S.K. Lamoreaux (Washington U., Seattle), 
       % P. Ageron, W. Mampe (Laue-Langevin Inst.), 
       % A. Steyerl (Munich, Tech. U.). 1990. 
         Phys. Lett. B {\bf 234}, 191 ( 1990 ).

\bibitem{QCD:theta}
         F.J. Yndurain, 
         {\it The theory of quark and gluon interactions\/} 
         ( Springer, Berlin ; New York, 1993 ).

         R.K. Bhaduri,
         {\it Models of the Nucleon\/}
         ( Addison-Wesley, Redwood, 1988 ).

\bibitem{QCD:CP}
       % THE STRONG CP PROBLEM REVISITED.
       % By Hai-Yang Cheng (Indiana U.). IUHET-125-REV., Feb 1987. 158pp. 
       % Revised version.
         Hai-Yang Cheng, 
         Phys. Rept. {\bf 158}, 1 ( 1988 ). 

\bibitem{FT:natural}
         G 't Hooft, in
         {\it  Recent Developments in Gauge Theories,\/}
         proceedings of NATO Advanced Study Institute, Cargese, France, 1979,
         edited by G 't Hooft {\it et al.\/}
       % C. Itzykson, (ed.), A. Jaffe, (ed.), H. Lehmann, (ed.),
       % P.K. Mitter, (ed.), I.M. Singer, (ed.), R. Stora, (ed.). 1980. 
         ( Plenum, New York, 1980 ).
       % ( Nato Advanced Study Institutes Series: Series B, Physics, 59 ). 

\bibitem{CP:A Nelson}
       % Naturally weak CP violation.
         A. Nelson,
         Phys. Lett. {\bf B136}, 387 ( 1984 ).

\bibitem{CP:PQ}
       % 1) CP CONSERVATION IN THE PRESENCE OF PSEUDOPARTICLES.
       % R.D. Peccei, H.R. Quinn (Stanford U., Phys. Dept.). 1990. 
       % In *Wolfenstein, L. (ed.): CP violation* 328-331. 
       % ( Phys. Rev. Lett. 38 (1977) 1440-1443 ) 
       % and Stanford Univ. - ITP-568 (77,rec.Apr.) 8 p.
       %
       % 2) CONSTRAINTS IMPOSED BY CP CONSERVATION IN THE 
       % PRESENCE OF INSTANTONS. 
         R.D. Peccei and H.R. Quinn, 
       % (Stanford U., ITP). ITP-572-STANFORD, May 1977. 
         Phys. Rev. D{\bf 16}, 1791 ( 1977 ); 
       % (* Title changed in journal *) 
       %
       % 3) CP CONSERVATION IN THE PRESENCE OF INSTANTONS.
       % R.D. Peccei and H. R. Quinn,
       % ITP-568-STANFORD, Mar 1977. 8pp. 
         Phys. Rev. Lett. {\bf 38}, 1440 ( 1977 ).
       % (* Title changed in journal *) 
       %
       % 4) SOME ASPECTS OF INSTANTONS.
       % By R.D. Peccei, Helen R. Quinn (Stanford U., ITP). 
       % ITP-555-STANFORD, Jan 1977. 35pp.
       % Published in Nuovo Cim.41A:309,1977 

\bibitem{QCD:C-T Chan}
         Chuan-Tsung Chan, Ph.D. thesis, University of Washington, 1996.

\bibitem{QCD:Fuji}
       % 1) PATH INTEGRAL MEASURE FOR GAUGE INVARIANT FERMION THEORIES.
         K. Fujikawa, 
       % ( Tokyo U., INS ). INS-328, Feb 1979. 10pp. 
         Phys. Rev. Lett. {\bf 42}, 1195 ( 1979 );
         {\it ibid.\/} {\bf 44}, 1733 ( 1980 );
       % 2) PATH INTEGRAL FOR GAUGE THEORIES WITH FERMIONS.
       % K. Fujikawa, 
       % ( Tokyo U., INS). INS-370, Jan 1980. 31pp. 
         Phys. Rev. D {\bf 21}, 2848 ( 1980 ); 
         {\it ibid.\/} {\bf 22}, 1499 ( E ) ( 1980 ).
       %
       % 3) COMMENT ON CHIRAL AND CONFORMAL ANOMALIES.
       % K. Fujikawa, 
       % ( SUNY, Stony Brook & Tokyo U., INS ). ITP-SB-80-12, 
       % ( Received Mar 1980 ). 11pp. 
       % Phys. Rev. Lett. {\bf 44}, 1733 ( 1980 ).

\bibitem{CP:Shabalin}
         E.P. Shabalin
       % Inst. of Theoretical & Experimental Phys., Moscow, USSR.
       % 1) Electric dipole moment of the neutron in gauge theory.
       % Uspekhi Fizicheskii Nauk.  vol.139, no.4.  pp. 561-85.  April 1983.
       % 94 refs.
       %
       % 2) THE ELECTRIC DIPOLE MOMENT OF THE NEUTRON IN A GAUGE THEORY.
       % E.P. Shabalin 
       % ( Moscow, ITEP ). ITEP-65-1982, 1982. 60pp. 
         Sov. Phys. Usp. {\bf 26}, 297 ( 1983 ); 
       %
       % 3) ELECTRIC DIPOLE MOMENT OF NEUTRON IN QUANTUM CHROMODYNAMICS.
       % (IN RUSSIAN).
       % By E.P. Shabalin (Moscow, ITEP). 1981. 
       % Published in Yad.Fiz.34:1598-1603,1981 
       % 15 refs.
         Sov. J. Nucl. Phys. {\bf 34}, ( 1981 )
       %
       % 4) Electric dipole moments of baryons in CP noninvariant
       % Kobayashi-Maskawa theory
       % Yadernaya Fizika.  vol.32, no.2.  pp. 443-7.  1980.
       % 11 refs.
       % Sov. J. Nucl. Phys. {\bf 32}, ( 1980 )
       %
       % 5) Electric dipole moment of a quark in the Kobayashi-Maskawa theory
       % with account of the gluonic corrections.
       % Yadernaya Fizika.  vol.31, no.6.  pp. 1665-79.  1980.
       % 9 refs.
       % Sov. J. Nucl. Phys. {\bf 31}, ( 1980 ).
       %
       % 6) THE U(1) PROBLEM, THE THETA TERM AND CP SYMMETRY VIOLATION. 
       % By E.P. Shabalin (Moscow, ITEP). 1982. 
       % Yadernaya Fizika.  vol.36, no.    pp. 981-985, 1982.
       % (Moscow, ITEP). ITEP-36-1982, 1982. 12pp. 
         {\it ibid.\/} {\bf 36}, 575 ( 1982 ).

\bibitem{NEDM:KW}
       % A SUM RULE CALCULATION OF THE NEUTRON ELECTRIC DIPOLE MOMENT
       % FROM A QUARK CHROMOELECTRIC DIPOLE COUPLING.
         I.I. Kogan and D. Wyler,
       % Ian. I. Kogan (British Columbia U.), Daniel Wyler (Zurich U.).
       % UBCTP-91-21, Aug 1991. 21pp. 
         Phys. Lett. B {\bf 274}, 100 ( 1992 ).

\bibitem{CP:SVZ}
       % CAN CONFINEMENT ENSURE NATURAL CP INVARIANCE OF STRONG
       % INTERACTIONS?
         M.A. Shifman, A.I. Vainshtein, and V.I. Zakharov, 
       % (Moscow, ITEP). ITEP-64-1979, 1979. 24pp. 
         Nucl. Phys. B {\bf 166}, 493 ( 1980 ).

\bibitem{CP:Aoki}
       % CALCULATING THE NEUTRON ELECTRIC DIPOLE MOMENT ON THE LATTICE.
       % By S. Aoki (SUNY, Stony Brook), A. Gocksch (Brookhaven),
       % A.V. Manohar (UC, San Diego), S.R. Sharpe (Washington U., Seattle). 
       % DOE-ER-40423-08-P90, Jun 1990. 6pp.
         S. Aoki, A. Gocksch, A.V. Manohar, and S.R. Sharpe, 
         Phys. Rev. Lett. {\bf 65}, 1092 ( 1990 ).

\bibitem{CP:A&H}
       % STRONG CP VIOLATION AND THE NEUTRON ELECTRIC DIPOLE MOMENT REVISITED.
       % By Sinya Aoki (SUNY, Stony Brook), Tetsuo Hatsuda (CERN). 
       % CERN-TH-5808-90, (Received Aug 1990). 27pp.
         S. Aoki and T. Hatsuda,
         Phys. Rev. D {\bf 45}, 2427 ( 1992 ).

\bibitem{CP:H-Y Cheng}
       % REANALYSIS OF STRONG CP VIOLATING EFFECTS IN 
       % CHIRAL PERTURBATION THEORY.
       % By Hai-Yang Cheng (Taiwan, Inst. Phys.). 
       % IP-ASTP-26-90, Dec 1990. 26pp. 
         Hai-Yang Cheng,
         Phys. Rev. D {\bf 44} 166 ( 1991 ).

\bibitem{UA1:tHooft}
       % HOW INSTANTONS SOLVE THE U(1) PROBLEM.
       % By G. 't Hooft (Utrecht U.). PRINT-86-0358 (UTRECHT), Apr 1986. 50pp. 
         G. 't Hooft,
         Phys. Rept. {\bf 142} 357 ( 1986 ).
       % (Also in *t'Hooft, G. (ed.): 
       % Under the spell of the gauge principle* 321-351) 

\bibitem{NEDM:C-T Chan}
       % Nucleon Electric Dipole Moments From QCD Sum Rules
         Chuan-Tsung Chan, E.M. Henley, and T. Meissner,
         UW-NUTH-Preprint    ( to be submitted for publication ).

\bibitem{QSR:SVZ}
       % 1) QCD AND RESONANCE PHYSICS. VECTOR NONET.
       % By M.A. Shifman, A.I. Vainshtein, V.I. Zakharov (Moscow, ITEP).
       % ITEP-94-1978, 1978. 64pp. 
         M.A. Shifman, A.I. Vainshtein, and V.I. Zakharov,
         Nucl. Phys. B {\bf 147}, 385, ( 1979 );
                       {\bf 147}, 448, ( 1979 );
                       {\bf 147}, 519, ( 1979 ).
       % (*ITEP-81-1978 and ITEP-94-1978 are combined in one journal article*)
       %
       % 2) QCD AND RESONANCE PHYSICS. NONPERTURBATIVE EFFECTS IN
       %    OPERATOR EXPANSION.
       % ITEP-73-1978, 1978. 68pp. 
       % Nucl. Phys. B {\bf 147}, 385, ( 1979 ) 
       % (*ITEP-73-1978 and ITEP-80-1978 are combined in one journal article*) 
       %  
       % 3) QCD AND RESONANCE PHYSICS. THE PI P A1 SYSTEM.
       % ITEP-81-1978, 1978. 48pp. 
       % Nucl. Phys. B {\bf 147}, 448, ( 1979 )
       % (*ITEP-81-1978 and ITEP-94-1978 are combined in one journal article*) 
       %
       % 4) QCD AND RESONANCE PHYSICS. SUM RULES.
       % ITEP-80-1978, 1978. 56pp. 
       % Nucl. Phys. B {\bf 147}, 385, ( 1979 )
       %  (*ITEP-73-1978 and ITEP-80-1978 are combined in one journal article*) 
       %
       % 5) QCD AND RESONANCE PHYSICS. THE RHO - OMEGA MIXING.
       % ITEP-99-1978, 1978. 27pp. 
       % Nucl. Phys. B {\bf 147}, 519, ( 1979 )

\bibitem{HQ: }
       % 1) HEAVY QUARK SYMMETRY.
       % By Nathan Isgur (CEBAF), Mark B. Wise (Cal Tech). 
       % CEBAF-TH-92-10, Aug 1991. 25pp.
       % Invited talk at Hadron 91, College Park, MD, Aug 12-16, 1991. 
       % In *Stone, S. (ed.): B physics* 158-209, 
       % In *Buras, A.J. (ed.), Lindner, M. (ed.): Heavy flavours* 234-285, 
       % and In *College Park 1991, Proceedings, Hadron '91* 549-572.
       % Southeast. Univ. R. A. Newport News 
       % - CEBAF-TH-92-10 (rec.Mar.92) 25 p. 
       %
       % 2) HEAVY QUARK SYMMETRY.
       % By Matthias Neubert (SLAC). SLAC-PUB-6263, Jun 1993. 186pp. 
       % M. Neubert,
       % Phys. Rept. {\bf 245}, 259 ( 1994 ). 
       % e-Print Archive: hep-ph/9306320
       % 
       % 3) PREASYMPTOTIC EFFECTS IN INCLUSIVE WEAK DECAYS OF CHARMED PARTICLES.
         M.A. Shifman and M.B. Voloshin, 
       % (Moscow, ITEP). ITEP-62-1984-mc (microfiche), 1984. 27pp.
         Sov. J. Nucl. Phys. {\bf 41} 120 ( 1985 );
       % Yad.Fiz.41:187-198,1985
       %
       % 4) LEPTON ENERGY DISTRIBUTIONS IN HEAVY MESON DECAYS FROM QCD.
       % By Junegone Chay, Howard Georgi, Benjamin Grinstein (Harvard U.).
       % HUTP-90-A035, Jun 1990. 9pp. 
         Junegone Chay, Howard Georgi, and Benjamin Grinstein,
         Phys. Lett. B {\bf 247}, 399 ( 1990 ).

%VIOLATION OF CP INVARIANCE, C ASYMMETRY, AND BARYON ASYMMETRY
%OF THE UNIVERSE.
%By A.D. Sakharov. 1967. 
%Published in Pisma Zh.Eksp.Teor.Fiz.5:32-35,1967, JETP Lett.5:24-27,1967,
%Sov.Phys.Usp.34:392-393,1991 (No.5) (Reprinted in *Kolb, E.W. (ed.), Turner,
%M.S. (ed.): The
%early universe* 371-373, and in *Lindley, D. (ed.) et al.: Cosmology and
%particle physics*
%106-109, and in Sov. Phys. Usp. 34 (1991) 392-393 [Usp. Fiz. Nauk 161 (1991)
%No. 5 61-64]) 
%Sakharov: Violation Of Cp Invariance, C Asymmetry, And Baryon Asy



   \end{references}
\end{document}